\documentclass{aastex63}

\usepackage{pythontex}
\usepackage{listings}
\usepackage{xcolor}
\usepackage[caption=false]{subfig}

\definecolor{codegreen}{rgb}{0,0.6,0}
\definecolor{codegray}{rgb}{0.5,0.5,0.5}
\definecolor{codepurple}{rgb}{0.58,0,0.82}
\definecolor{backcolour}{rgb}{0.95,0.95,0.92}

\lstdefinestyle{mystyle}{
    backgroundcolor=\color{backcolour},   
    commentstyle=\color{codegreen},
    keywordstyle=\color{magenta},
    numberstyle=\tiny\color{codegray},
    stringstyle=\color{codepurple},
    basicstyle=\ttfamily\footnotesize,
    breakatwhitespace=false,         
    breaklines=true,                 
    captionpos=b,                    
    keepspaces=true,                 
    numbers=left,                    
    numbersep=5pt,                  
    showspaces=false,                
    showstringspaces=false,
    showtabs=false,                  
    tabsize=2
}

\newcommand{\plc}{PyLightcurve}
\newcommand{\plct}{PyLightcurve-torch}
\newcommand{\torch}{PyTorch}
\newcommand{\np}{NumPy}

\submitjournal{PASP}

\shorttitle{\plct{}}
\shortauthors{Morvan et al.}

\graphicspath{{./}{}}

\begin{document}

\title{\plct{}: a transit modelling package for deep learning applications in \torch{}}

\correspondingauthor{Mario Morvan}
\email{mario.morvan.18@ucl.ac.uk}

\author[0000-0001-8587-2112]{Mario Morvan}
\affiliation{Department of Physics and Astronomy, University College London, Gower Street, London, WC1E 6BT, UK}
\author[0000-0003-3840-1793]{Angelos Tsiaras}
\affiliation{Department of Physics and Astronomy, University College London, Gower Street, London, WC1E 6BT, UK}
\author[0000-0001-8453-7574]{Nikolaos Nikolaou}
\affiliation{Department of Physics and Astronomy, University College London, Gower Street, London, WC1E 6BT, UK}
\author[0000-0002-4205-5267]{Ingo P. Waldmann}
\affiliation{Department of Physics and Astronomy, University College London, Gower Street, London, WC1E 6BT, UK}




\begin{abstract}
We present a new open source python package\footnote{\plct{} can be found on the following 
Github repository: https://github.com/ucl-exoplanets/pylightcurve-torch}, based on \plc{} \citep{tsiaras_pylightcurve_nodate} and \torch{} \citep{paszke_pytorch_2019}, tailored for efficient computation and automatic differentiation of exoplanetary transits. The classes and functions implemented are fully vectorised, natively GPU-compatible and differentiable with respect to the stellar and planetary parameters. This makes \plct{} suitable for traditional forward computation of transits, but also extends the range of possible applications with inference and optimisation algorithms requiring access to the gradients of the physical model. This endeavour is aimed at fostering the use of deep learning in exoplanets research, motivated by an ever increasing amount of stellar light curves data and various incentives for the improvement of detection and characterisation techniques. 

\end{abstract}


\section{Introduction} \label{sec:intro}
Exoplanets science discoveries have relied largely on our ability to extract precise information from stellar light curves. In the case of transiting exoplanets, this requires high precision photometric or spectroscopic measurements, and involves a transit model with one or several parameters to be determined. However, transit light curves often contain other sources of temporal variability caused by the instrument or the host star, which need to be accounted for in a preliminary or joint processing step. Consequently, forward modelling of transits needs to be considered as part of the data processing pipeline yielding the planetary parameters. 

The complexity and growing amount of exoplanets light curves hint at the use of deep learning to help alleviating the issues encountered with traditional modelling techniques. Indeed, tremendous progress has been made recently in time series analysis owing to the recent development of deep learning, producing several successful applications and promising solutions for problems ranging from time series classification \citep{fawaz_deep_2019} and forecasting \citep{hewamalage_recurrent_2020} to denoising and anomaly detection \cite{chalapathy_deep_2019}. Amongst the various fields benefiting from such technical progress, astronomy has not been an exception. Several recent contributions to exoplanetary science (\citealp{zingales_exogan:_2018}, \citealp{passegger_carmenes_2020}, \citealp{yip_peeking_nodate} for spectra, \citealp{shallue_identifying_2018}, \citealp{pearson_searching_2018}, \citealp{morvan_detrending_2020} and \citealp{nikolaou_the_ariel_ml_challenge_nodate} for photometric light curves) involve the use of deep learning methods.

The recent successes of deep learning can be traced back to the introduction of backpropagation as a technique which has allowed for the efficient optimisation of neural networks \citep{rumelhart_learning_1986}. Today all major deep learning frameworks including TensorFlow \citep{abadi_tensorflow_2016} and Pytorch implement a way to automatically compute gradients of scalar outputs of functions, with respect to their inputs and parameters. To use and include a function in a data flow graph created by one of the deep frameworks above, the function must first be implemented using the buildings blocks of the framework: the functions and the numerical objects (multidimensional arrays, often called \emph{tensors}) specific to the framework's language. So far, to the best of the authors' knowledge, none of the existing transit modelling codes has been designed to allow automatic differentiation and thereby joint end-to-end training with artificial neural networks. This is precisely the purpose of \plct{}, which provides a user-friendly transit modelling tool to facilitate the generation, inference and optimisation of transit models in a framework compatible with deep learning modules. We chose the language of \torch{} due to its user-friendliness, flexibility, efficiency and growing popularity. As the \torch{} syntax is very close to that of \np{}, it facilitates the easy conversion of \np{} codes to \torch{}, and reduces the learning curve for research communities who are used to scientific programming in \np{}. 

\plct{} is adapted from \plc{}\footnote{\url{https://github.com/ucl-exoplanets/pylightcurve}}, which is one of the most efficient open-source transit modelling packages available. \plc{} performs some numerical approximations rather than solving the fully analytical transit model to enable vectorisation of computations with \np{} \citep{harris_array_2020} and gain in efficiency, thus providing a good template for designing differentiable and scalable code. Four different limb-darkening laws are natively available in \plc{}, as well as several utilities for database access and fitting that we are not considering here. For more details about \plc{}'s physics models, implementation and performance, see \cite{tsiaras_pylightcurve_nodate}. 


The remainder of this article discusses the implementation (Section \ref{sec:code}), performance (Section \ref{sec:perf}) and applicability (Section \ref{sec:applis}) of \plct{}.

    \section{Code design} \label{sec:code}

The numerical programming code  was adapted from \plc{} transit modelling library \cite{tsiaras_pylightcurve_nodate}. Indeed the main functions \verb+exoplanet_orbit, transit_duration, transit_flux_drop, transit and eclipse+ have been translated to \torch{} while preserving their names, structure, and parameters. However several major changes have been introduced along with the conversion to PyTorch. These have been summarised in the list below.

\paragraph{\textbf{From \np{} to \torch{}}} \np{} arrays and operations have been converted respectively to \torch{} tensors and their corresponding operations. This means that the input parameters of corresponding main functions must now be of type \verb+torch.tensor+ and of shape broadcastable to \verb+(batch_size,)+ where \verb+batch_size+ is the number of instances of each transit parameter. While \torch{} tensors share many similarities with \np{} arrays, they further allow for \textbf{GPU acceleration} and \textbf{automatic differentiation}. 

\paragraph{\textbf{Vectorisation}} Main functions have been further vectorised to allow inputting 1D-arrays for each transit parameter in addition to scalars. Expressed in another way, this allows the user to provide batches of inputs to the main functions instead of individual sets of parameters. In the \torch{} version of \plc{}, this not only enables batch learning - i.e. optimisation based on groups of observations considered jointly - but also fully leverages the GPU acceleration advantages on multi-dimensional tensors. 

\paragraph{\textbf{Flexibility of input shapes}} By allowing inputs of parameters of broadcastable shapes, the use of the main functions remains intuitively flexible, while saving memory and time in some specific cases. Indeed, when intermediate computations can be shared across a batch, such as the planetary positions vector for scalar orbital parameters, only a vector of batch dimension 1 is computed and used for later computations of transit flux drops, even if the latter are multidimensional.

\paragraph{\textbf{\texttt{TransitModule} class}} A class named \verb+TransitModule+ has been implemented to facilitate the use of \verb+transit+ and \verb+eclipse+ functions, their optimisation and embedding in deep learning pipelines. Indeed \verb+TransitModule+ first manages transit parameters and intermediate computations in an object-oriented fashion (see Listing \ref{code_example} for a basic example). For convenience, the transit parameters passed as attributes of a module undergo checks, type, shape and device casting to make sure correct inputs are passed to the \torch{} functions performing the actual transit computations. Secondly, \verb+TransitModule+ inherits the \verb+torch.nn.Module+ along with its methods and parameters internal management. Furthermore, as the main parent class of all neural networks implemented in \torch{}, instances of \verb+torch.nn.Module+ can easily be combined together, facilitating the embedding or combination of our transit models with neural networks. 







\begin{lstlisting}[language=Python, label=code_example, caption=Basic use of wrapper class \texttt{TransitModule} computing transit and/or eclipse flux while inheriting \texttt{torch.Module} class.]
from pylightcurve_torch import TransitModule
...

# Model definition
tm = TransitModule(time, **transit_params)  # transit_params is a dict of parameters 
tm.activate_gradient('rp')                  # Gradient activation for parameter 'rp'

# Forward transit computation
flux = tm()                                 # with module's defined parameters
flux = tm(e=0., t0=3.4)                     # with substituted external parameters

# Loss and backward pass
err = loss(flux, **data)                    # loss computation (PyTorch function with scalar output) 
err.backward()                              # backward propagation of gradients 
tm.rp.grad                                  # accessing the computed gradient w.r.t. parameter 'rp'

\end{lstlisting}

\section{Performance} \label{sec:perf}

Several tests were conducted to assess the performance of \plct{} and compare it to \plc{}. First of all, a sanity check was carried out to ensure both codes provide the same outputs when provided with the same inputs, up to to a precision level below $0.1 ppm$, on average, for default precision settings.

\begin{figure*}[ht]
\begin{center}
\includegraphics[width=0.8\textwidth]{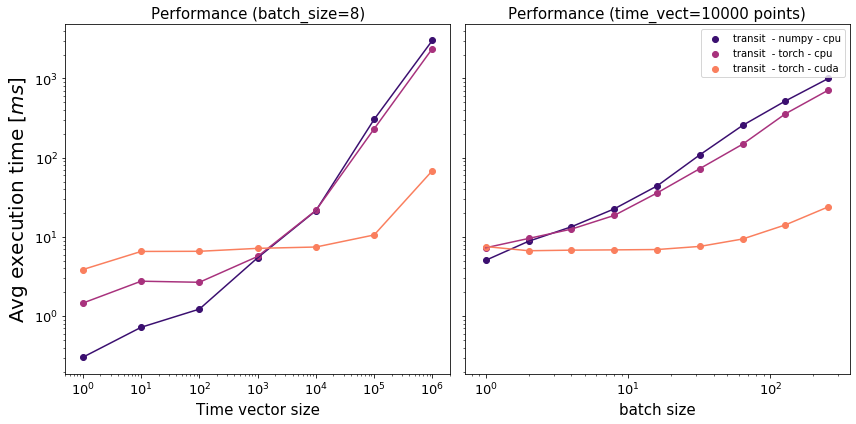}
\caption{Execution times of \plc{}  \texttt{transit} functions in \np{} and \torch{}, averaged over 50 executions. \torch{} functions have been run onto a CPU cluster (purple curve) and a GPU core (orange curve). \textbf{Left: } batch size fixed to $8$ while increasing the length of the input time series. \textbf{Right: } the input time series size was fixed to $10000$ points while increasing the batch size. \label{fig:perf}}
\end{center}
\end{figure*}

Two experiments were then performed with the aim of comparing the computational efficiency of both codes on CPU and GPU machines. In the first case the time array/tensor length was varied while keeping fixed the number of transit parameters, and in the second case the time array/tensor was fixed to 1000 while the number of parameters varied. The results are presented in Figure \ref{fig:perf}, suggesting very similar performances for the \np{} and the torch-cpu versions of \plc{}  \verb+transit+ function. However, the GPU runs show a significant reduction in computation time, which is no longer increasing linearly with the input sizes but rather plateauing under $\sim 10ms$ for time tensor sizes smaller than $10
^6$ and batch\_size smaller than $\sim 256$. Although these specific thresholds depend on the architecture used - which in our case consisted of 10 CPU cores, 70GB memory and 1 Tesla-V100 GPU core -  we expect GPUs to bring significant improvements in most configurations and use cases. 

The ability to handle large datasets and maintain efficiency while computing a transit function or its gradient opens up new possibilities for processing exoplanet light curve datasets. Furthermore, automatic differentiability now allows the \verb+transit+ function or \verb+TransitModule+ to be included dynamically in deep learning pipelines and involved in their end-to-end optimisation. The next section discusses the potential for new applications brought by  \plct{}. 

\newpage
\section{Applications} \label{sec:applis}
This section, rather than being an exhaustive list of applications, aims to provide a quick overview of the types of novel applications afforded by the use of \plct{} for designing both generative and discriminative models. 

\subsection{Generation}
\plct{}  can be used in a static generative mode, to efficiently simulate primary and secondary transit light curves. In this case the gradients may not need to be activated (static mode: \verb+requires_grad=False+), depending on the problem considered. Indeed, one can, for instance, create artificial datasets statically and use them to train models for various problems such as transit depth regression or event classification. While the simulated transits, created as \torch{} tensors, enable GPU acceleration, the conversion to \np{} arrays is still possible and very cheap computationally, simply by means of calling the \verb+.numpy()+ method available for \verb+torch.Tensor+ objects.

\subsection{Gradient-based optimisation} 

Having an efficient access to the transit model's output gradient with respect to the transit parameters enables gradient-based optimisation without having to use approximate methods to compute gradients. Since the reverse-mode automatic differentiation (i.e. backpropagation) computes gradients efficiently, particularly for scalar outputs, we are led to consider the problem of scalar loss function minimisation, an integral aspect of machine learning. Any loss function can be computed from the transit model output using differentiable functions to allow backward gradient-optimisation. \torch{} implements a number of off-the-shelf optimisers which make use of the first or higher order derivatives of parameters. Moreover, several popular MCMC sampling algorithms such as Hamiltonian Monte Carlo and NUTS (ref) also require the availability of gradients, hence opening up the possibility for this class of MCMC algorithms to be used to derive posterior distributions for our transit models. Implementations of HMC and NUTS are available in the probabilistic programming language Pyro  \citep{bingham_pyro_2018}, which is also based on Python and \torch{}. Besides exact bayesian inference with MCMC, it is worth noting that Pyro also provides a convenient framework for probabilistic, flexible and deep inference of parameters, and has been designed to be fully compatible with \torch{} tensors, functions and modules. 

\subsection{Combination with neural networks} 
\begin{figure}
    \centering 
    \subfloat[\centering Neural net. in input]{
      \centering\includegraphics[width=0.15\textwidth]{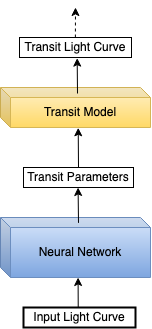}
      \label{fig:block_input}
    }\qquad
    \subfloat[\footnotesize  Neural net. in parallel]{
      \centering\includegraphics[width=0.3\textwidth]{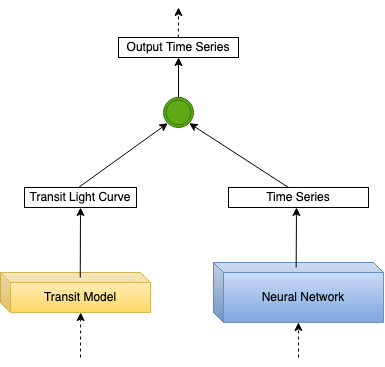}
      \label{fig:block_par}
    }\qquad
    \subfloat[\footnotesize  Neural net. in output]{
        \centering\includegraphics[width=0.17\textwidth]{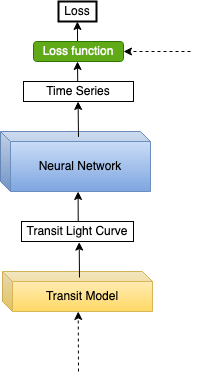}
        \label{fig:block_output}
    }
\caption{Schematic building blocks showing examples of combination between a neural network and differentiable transit model inputs (a) and/or outputs (b and c). Dotted arrows indicate possible additional inputs and outputs. The circular green node represents here a non parametric operation such as item-wise multiplication or addition. The loss function (in green) computes a scalar value from a time series in this case, optionally using some additional scalar or tensor values.}    
\end{figure}

The flexibility afforded by the autograd package and torch modules makes it particularly easy to connect the input and the output of our transit model with other differentiable functions and modules. All parameters can then be optimised in an end-to-end mode through the computational graph automatically built when defining and operating on the tensors. In practice, this means that any other differentiable module or function can:
\begin{itemize}
    \item provide the transit parameters as \textbf{input} of the transit model. A schematic example of this setup is shown in Figure \ref{fig:block_input}.
    \item be used in \textbf{parallel} with the transit model to provide other scalars, vectors or time series to be combined subsequently with the transit output. This setup is, for example, suitable for time series decomposition of transit light curves by means of generative models. A schematic example of this setup is shown in Figure \ref{fig:block_par}.
    \item be applied to the \textbf{output} time series of the transit model, transforming it to another time-series, a vector or a scalar. Note that to perform gradient-based optimisation, a loss function outputting a scalar value will need to be at the end of any functional flow. A schematic example of this setup is shown on Figure \ref{fig:block_output}.
\end{itemize}

\subsection{Experiment}

Let us consider a simple example of regression problem where we use the transit module as an extra term in the objective loss function of a neural network. As such, this experiment falls into the case (a) discussed above, where a neural network output is used as input to a transit model. In the present experiment, the main task is to build a predictor for $y = R_P/R_s$, assuming the other transit parameters, $\theta$, to be known. The dataset is composed of $2000$ light curves generated by \plct{} with added Gaussian noise. All the lightcurves are univariate time-series with $T=1000$ uniform timesteps and parameters fixed except for the target $R_P/R_S$ and the inclination $i$ which are sampled from uniform distributions. We denote by $Y$ the set of ground-truth targets and by $\mathcal{M}$ the transit model which generated the light curves.

Two identical neural network models $m_1$ and $m_2$ are trained on this regression problem using \emph{different loss functions}. Indeed, $m_1$ is only trained with a mean-squared error loss on the target parameters:
$$\mathcal{L}_1 = \frac{1}{|Y|}\sum_{y \in Y}{(y - \hat{y})^2} \text{~,}$$
whereas $m_2$ is also trained to reconstruct the transit flux by embedding the transit model as an additional term in the loss function: 
$$\mathcal{L}_2 = \lambda \mathcal{L}_{regression} + (1-\lambda)\mathcal{L}_{reconstruction} $$
where :
\begin{itemize}
    \item $\mathcal{L}_{regression}$ is identical to $\mathcal{L}_1$ but with predicted values $\hat{y}$ from $m_2$ instead. 
    \item The second term is a \textbf{transit reconstruction loss}, measured as the mean-squared error between the input light curve and the reconstructed transit light curve: $$\mathcal{L}_{reconstruction} = \frac{1}{|Y|T}\sum_{y \in Y}{\sum_{t \in {1..T}}{ (\mathcal{M}(\hat{y}, \theta)_t - x_t)^2}}$$
    \item $\lambda$ is a scalar hyperparameter balancing the relative importance between the two loss terms.
\end{itemize}

The neural network architecture chosen consists of 4 convolutional blocks followed by 2 linear layers. Both models are trained using the Adam optimiser for 20000 steps. Even though the networks and $\lambda$ parameters would be suited for hyperparameter optimisation, finding the optimal solution to this problem would go beyond the scope of this study, and here we simply present the results for a comparison of cases $\lambda = 0$ ($\mathcal{L}_2 = \mathcal{L}_{regression} = \mathcal{L}_1) $ and $\lambda=0.5$ (equal contributions from $\mathcal{L}_{regression}$ and $\mathcal{L}_{reconstruction}$)  .      

\begin{figure*}[ht]
\begin{center}
\includegraphics[width=0.8\textwidth]{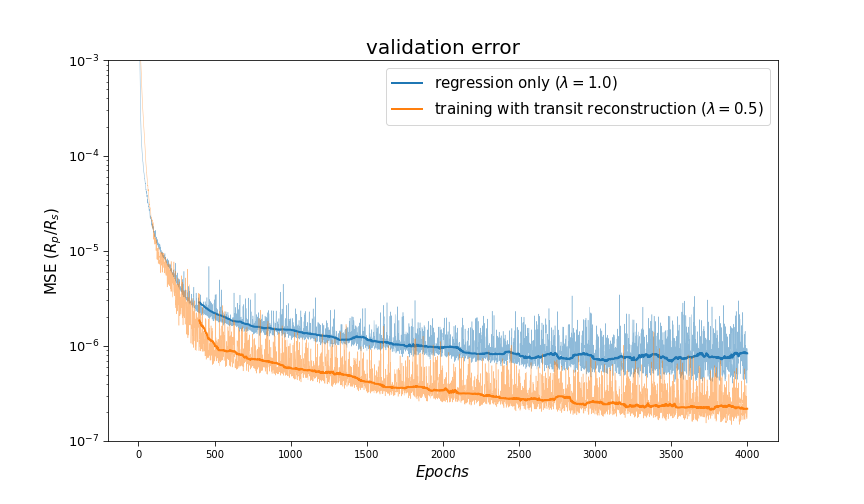}
\caption{ Validation mean-squared errors for $R_P/R_S$ prediction with (orange curve) and without (blue curve) including a transit reconstruction term in the traning loss function. The thicker orange and blue curves merely show the moving median of the respective validation errors with a window size of 100 steps.\label{fig:exp}}
\end{center}
\end{figure*}

The performance of both models is measured with the mean-squared-error of the predicted transit depths. The evolution of this metric evaluated on a validation subset of 400 lightcurves during learning is presented in Figure \ref{fig:exp}, which shows a significantly lower validation error for the model $m_2$ trained with the transit reconstruction term in the loss function. Indeed this model reaches $m_1$'s final performance far before $m_1$ (in only about 2000 epochs) and continues to further improve its performance until the maximum number of epochs (20000) is reached. This indicates a clear advantage given to the model informed from the transit function and jointly trained with the proxy task of reconstructing the transit shape.

\newpage
\section{Summary and Discussion}

In this article we outlined several possible uses of \plct{}  for generating and optimising transits as well as embedding them in machine learning flows optimised end-to-end by gradient descent. While these examples are aimed at suggesting basic use-cases, the list is not exhaustive and various other uses might be designed in the future. Furthermore, the rich ecosystem of open-source libraries building up in \torch{} should provide ideas and help in the development of more complex applications, making use of existing support - e.g. for deep probabilistic programming, deep Gaussian processes, Bayesian hyperparameter optimisation, gradient boosting and various implemented neural networks architectures. 

By providing a differentiable and GPU-accelerated transit code, \plct{}  aims to facilitate and widen the use of deep learning in exoplanets research. It bridges the gap between the precision of physical transit models and the scalability of neural networks, allowing for the efficient modelling of thousands of transit light curves now commonly available with exoplanets transit surveys such as Kepler \citep{borucki_kepler_2010} and TESS \citep{ricker_transiting_2015} from space or HATNet \citep{bakos_wide-field_2004}, SuperWASP \citep{pollacco_wasp_2006} and NGTS \citep{wheatley_next_2018} from the ground. Conversely, we hope that this code will also make exoplanets science more accessible to the machine learning community and more generally inspire the development of physics-based deep learning applications. 


\acknowledgments
This project has received funding from the European Research Council (ERC) under the European Union's Horizon 2020 research and innovation programme (grant agreement No 758892, ExoAI) and the European Union's Horizon 2020 COMPET programme (grant agreement No 776403, ExoplANETS A). Furthermore, we acknowledge funding by the Science and Technology Funding Council (STFC) grants: ST/K502406/1, ST/P000282/1, ST/P002153/1 and ST/S002634/1. We are grateful for the support of the NVIDIA Corporation through the NVIDIA GPU Grant program.

\bibliography{main.bib}{}
\bibliographystyle{aasjournal}



\end{document}